\documentclass[aps, prd, preprint, amsfonts,
  amssymb, amsmath, showpacs, a4paper,nofootinbib,superscriptaddress]{revtex4-1}  
\usepackage[T1]{fontenc} 

\usepackage{slashed}
\usepackage{braket}

\newcommand{\be}{\begin{equation}}
\newcommand{\ee}{\end{equation}}
\newcommand{\bea}{\begin{eqnarray}}
\newcommand{\eea}{\end{eqnarray}}

\newcommand{\im}{\text{Im}}

\usepackage{color}

\begin{document} 

\title{Gauge invariance and the Englert-Brout-Higgs mechanism in\\ non-Hermitian field theories}

\author{Jean Alexandre}
\email{jean.alexandre@kcl.ac.uk}
\affiliation{Department of Physics, King's College London,\\ 
London WC2R 2LS, United Kingdom}

\author{John Ellis}
\email{john.ellis@cern.ch}
\affiliation{Department of Physics, King's College London,\\ 
London WC2R 2LS, United Kingdom}
\affiliation{National Institute of Chemical Physics \& Biophysics, R\"avala 10, 10143 Tallinn, Estonia}
\affiliation{Theoretical Physics Department, CERN, CH-1211 Geneva 23, Switzerland}

\author{Peter Millington}
\email{p.millington@nottingham.ac.uk}
\affiliation{School of Physics and Astronomy, University of Nottingham,\\ Nottingham NG7 2RD, United Kingdom\vspace{1em}}

\author{Dries Seynaeve}
\email{dries.seynaeve@kcl.ac.uk}
\affiliation{Department of Physics, King's College London,\\ 
London WC2R 2LS, United Kingdom}

\begin{abstract}

We discuss $\mathcal{PT}$-symmetric Abelian gauge field theories, as well as
their extension to the Englert-Brout-Higgs mechanism for generating
a mass for a vector boson. Gauge invariance is not straightforward, and we discuss the different
related problems, as well as a solution which consists in coupling the gauge field to a current that is not conserved.
Non-Hermiticity then necessarily precludes the Lorenz gauge condition but nevertheless allows for a consistent
formulation of the theory. We therefore generalise the Englert-Brout-Higgs mechanism
to $\mathcal{PT}$-symmetric field theories, opening the way to constructing non-Hermitian
extensions of the Standard Model and new scenarios for particle model-building.\\
~~\\
KCL-PH-TH/2018-40, CERN-TH/2018-183
~~\\
April 2019

\end{abstract}

\maketitle


\section{Introduction}

There has been much work in recent years on quantum-mechanical models with non-Hermitian, $\mathcal{PT}$-symmetric
Hamiltonians~\cite{Bender:1998ke,Bender:2002vv,Bender:2005tb}, which have become an important area of research in integrated photonics 
and other fields~\cite{Longhi,El-Ganainy,Ashida}  --- see Ref.~\cite{Znojil:2017viu} for a review of relations to conventional models with Hermitian Hamiltonians. 
Quantum field theories (QFTs) with non-Hermitian Hamiltonians have also attracted interest, including a model with an $i\phi^3$ 
scalar interaction~\cite{Blencowe:1997sy,Bender:2004vn,Jones:2004gp,Bender:2013qp,Shalaby:2017wux}, which
was shown in the framework of $\mathcal{PT}$-symmetric QFT to have a physically meaningful effective potential 
despite its being unbounded from below~\cite{Bender:2015uxa}, and a $\mathcal{PT}$-symmetric $-\phi^4$ model featuring asymptotic freedom \cite{Shalaby:2009xda}.
A $\mathcal{PT}$-symmetric theory with a non-Hermitian fermion mass term 
$\mu \bar\psi\gamma^5\psi$ was considered in Ref.~\cite{Bender:2005hf}, and it was shown in Ref.~\cite{Alexandre:2015oha}
that this model possesses a conserved current and that its $\mathcal{PT}$ symmetry is consistent with unitarity. 

Among applications to particle physics, the possibility of using the non-Hermitian term {$\mu \bar\psi\gamma^5\psi$} to describe neutrino masses was considered
in Refs.~\cite{JonesSmith:2009wy,Alexandre:2015kra,Alexandre:2017fpq}, and the application of non-Hermitian QFT to 
neutrino oscillations was considered in Ref.~\cite{Ohlsson:2015xsa}. 
A lattice version of a non-Hermitian fermionic model was studied in Ref.~\cite{Chernodub:2017lmx}, where it was shown that
this model could accommodate different numbers of left-handed and right-handed excitations, consistent with the fermionic 
symmetry current found in Ref.~\cite{Alexandre:2015oha}. There have also been applications of non-Hermitian QFT
to dark matter~\cite{Rodionov:2017dqt} and to decays of the Higgs boson~\cite{Korchin:2016rsf}, 
and it was argued in Ref.~\cite{Raval:2018kqg} that the $\mathcal{PT}$-symmetry properties of ghost fields are
relevant for the confinement phase transition in QCD. Effective non-Hermitian Hamiltonians with complex spectra
are also known to play a role in the description of unstable systems with particle mixing (see, e.g., Ref.~\cite{Pilaftsis:1997dr}). 

In this work, we show how the gauge symmetries of non-Hermitian and $\mathcal{PT}$-symmetric theories may be broken
via a generalisation of the Englert-Brout-Higgs mechanism~\cite{Englert:1964et,Higgs:1964pj}, opening the way to significant extensions of the Standard Model
and other particle physics theories.

This extension is non-trivial, {as it} was discovered in Ref.~\cite{Alexandre:2017foi} (for a summary, see Ref.~\cite{Alexandre:2017erl}) that the existence of a conserved
current in a $\mathcal{PT}$-symmetric QFT does {\it not} correspond to a symmetry of the Lagrangian ${\cal L}$.
Instead, it corresponds to a non-trivial transformation of the non-Hermitian part of ${\cal L}$, thereby
evading Noether's theorem~\cite{Noether}, in that symmetries of a $\mathcal{PT}$-symmetric 
Lagrangian are {\it not} related to conserved currents. We emphasise that conserved currents do exist though, as in the Hermitian case.  
This striking observation raised the interesting question 
whether $\mathcal{PT}$-symmetric QFTs exhibit an analogue of the spontaneous breaking of a global symmetry
that is familiar in Hermitian QFTs and, if so, whether this spontaneous symmetry breaking is accompanied by a 
massless Goldstone mode~\cite{GSW1,GSW2,GSW3}. 

The answers to both questions are {\it yes}~\cite{AEMS1}. One can define consistently a saddle point of the potential in a
$\mathcal{PT}$-symmetric QFT with a quartic scalar potential in which the scalar fields have 
symmetry-breaking vacuum expectation values (vev's) that are accompanied by a massless Goldstone mode. The existence of the latter follows from current conservation, 
even though the Lagrangian is not invariant under the corresponding
field transformations. The existence of this Goldstone mode was confirmed by an explicit calculation of
the effective potential at the tree and one-loop levels.
Our analysis of these questions was based on a formulation of a
non-Hermitian QFT that included a consistent quantisation of the path integral. This is possible
because the $\mathcal{PT}$-symmetric theory possesses a complete set of real energy eigenstates, 
which allow for saddle points about which the integration of quantum fluctuations is well-defined. 
The conventional quantisation of
the path integral for a Hermitian scalar Lagrangian can be extended consistently to the
non-Hermitian case by using $\mathcal{PT}$ conjugation instead of Hermitian conjugation~\cite{AEMS1}.

These developments have opened the way to exploring whether
the Englert-Brout-Higgs mechanism \cite{Englert:1964et,Higgs:1964pj} for generating masses for
gauge bosons also has a generalisation to the non-Hermitian case. As we show in this paper, the answer
is again {\it yes}. This might seem surprising, since coupling the gauge field to the conserved current does not lead to 
a gauge-invariant Lagrangian. However, we show how a consistent model can be obtained when coupling the gauge field to a non-conserved current,
provided a covariant gauge fixing term is present in the Lagrangian.

The layout of this paper is as follows. In Sec.~\ref{sec:symm}, we begin by setting up the $\mathcal{PT}$-symmetric QFT that
we use for our analysis. After reviewing symmetries and conservation laws in this context, we then discuss spontaneous symmetry 
breaking and the Goldstone mode in this theory in Sec.~\ref{sec:global}. The gauging of this
 $\mathcal{PT}$-symmetric model is described in Sec.~\ref{sec:gauged}, and the associated Englert-Brout-Higgs mechanism in Sec.~\ref{sec:gauge}. 
Finally, we summarise our conclusions and discuss perspectives for possible future research in Sec.~\ref{sec:conx}.

\section{Symmetries and conservation laws}
\label{sec:symm}

We start by considering a theory with two complex scalar fields $\phi_{1,2}$ described by the Lagrangian density first studied in Refs.~\cite{Alexandre:2017foi, Alexandre:2017erl}
\be \label{eq:LagrangianFreeScalar}
\mathcal{L}\  =\ \partial_\alpha \phi_1^\star \partial^\alpha \phi_1 + \partial_\alpha \phi_2^\star \partial^\alpha \phi_2 - m_1^2 |\phi_1|^2 - m_2^2 |\phi_2|^2 
-\mu^2\big( \phi_1^\star \phi_2 -\phi_2^\star \phi_1\big)~,
\ee
whose squared mass eigenvalues are given by
\be
M_\pm^2=\frac{1}{2}(m_1^2+m_2^2)\pm\frac{1}{2}\sqrt{(m_1^2-m_2^2)^2-4\mu^4}~.
\ee
These are real as long as
\be\label{eta}
\eta\equiv\frac{2\mu^2}{|m_1^2-m_2^2|}\le1~.
\ee
The Lagrangian (\ref{eq:LagrangianFreeScalar}) is left invariant by the $\mathcal{PT}$ transformation
\be 
\mathcal{PT}: ~~~~\Phi=\begin{pmatrix}
\phi_1 \\ \phi_2
\end{pmatrix} \ \rightarrow\ \begin{pmatrix}
\phi_1^\star \\ -\phi_2^\star
\end{pmatrix}~.
\ee
The field $\phi_1$ transforms as a scalar under parity, i.e.~$\mathcal{P}:\phi_1\to +\,\phi_1$, and the field $\phi_2$ transforms as a pseudoscalar, i.e.~$\mathcal{P}:\phi_2\to -\,\phi_2$. 
Time-reversal $\mathcal{T}$ is taken to be the usual anti-linear operator, whose action is equivalent to complex conjugation on the c-number fields $\phi_1$ and $\phi_2$. 
(We do not consider the discrete symmetries of this theory in Fock space.)

Since the Lagrangian (\ref{eq:LagrangianFreeScalar}) is not Hermitian, the corresponding action $S$ has the imaginary part
\be\label{imaginary}
{\rm Im}\,S = i\mu^2\int {\rm d}^4x\big( \phi_1^\star \phi_2 -\phi_2^\star \phi_1\big)~,
\ee
implying that the following equations of motion are not equivalent:
\be
\frac{\delta S}{\delta\Phi^\dagger}\equiv\frac{\partial \mathcal{L}}{\partial \Phi^{\dagger}}-\partial_{\alpha}\,\frac{\partial \mathcal{L}}{\partial (\partial_{\alpha}\Phi^{\dagger})} 
= 0~~~~\nLeftrightarrow~~~~\frac{\delta S}{\delta\Phi}\equiv\frac{\partial \mathcal{L}}{\partial \Phi}-\partial_{\alpha}\,\frac{\partial \mathcal{L}}{\partial (\partial_{\alpha}\Phi)}=0~.
\ee
(We emphasise that the functional variation $\delta S/\delta \Phi^{(\dagger)}$ is understood here and in what follows as a shorthand for the ``naive'' variation 
of the action that yields the usual Euler-Lagrange equations.) It would appear, therefore, that there is some ambiguity in the definition of the equations of motion.
This ambiguity can be resolved, however, by carefully defining the states (and their inner product) and considering the variational  
procedure in detail \cite{Alexandre:2017foi, Alexandre:2017erl}, as we outline below. 

If we neglect surface terms, we can write the Lagrangian (\ref{eq:LagrangianFreeScalar}) in the symmetric form 
\be
\label{eq:Lsym}
\mathcal{L}=\Phi^\ddagger\begin{pmatrix} -\Box-m_1^2 & -\mu^2 \\ -\mu^2 & \Box+m_2^2 \end{pmatrix}\Phi~,
\ee
where $\Phi^\ddagger=(\phi_1^\star~,~-\phi_2^\star)$, which shows that the conjugate variables (and states) to consider here are the $\mathcal{PT}$-conjugate fields $\{\Phi,\Phi^\ddagger\}$, 
instead of the Hermitian-conjugate fields $\{\Phi,\Phi^\dagger\}$. It nevertheless remains the case that we have a choice to define the equations of motion by varying 
Eq.~\eqref{eq:Lsym} with respect to $\Phi^{\ddagger}$ or $\Phi$. Taking the former variation, the equations of motion are given by
\be\label{equamotinit}
\frac{\delta S}{\delta\Phi^\ddagger}\equiv\frac{\partial \mathcal{L}}{\partial \Phi^{\ddagger}}-\partial_{\alpha}\,\frac{\partial \mathcal{L}}{\partial (\partial_{\alpha}\Phi^{\ddagger})}
= 0~~~~\mbox{and}~~~~\left(\frac{\delta S}{\delta\Phi}\right)^\ddagger\equiv\bigg(\frac{\partial \mathcal{L}}{\partial \Phi}-\partial_{\alpha}\,\frac{\partial \mathcal{L}}{\partial (\partial_{\alpha}\Phi)}\bigg)^{\ddagger}=0~.
\ee
This implies, however, that
\be
\label{eq:deltaSdeltaPhi}
\frac{\delta S}{\delta\Phi}\equiv \frac{\partial \mathcal{L}}{\partial \Phi}-\partial_{\alpha}\,\frac{\partial \mathcal{L}}{\partial (\partial_{\alpha}\Phi)}\ne0~,
\ee
except when we have the trivial solution $\phi_1=\phi_2=0$. For non-trivial solutions, the non-vanishing of the complementary variation in Eq.~\eqref{eq:deltaSdeltaPhi} is necessarily 
supported by non-vanishing surface terms or external sources, as explained in detail in references~\cite{Alexandre:2017foi, Alexandre:2017erl, AEMS1}.

The equations of motion defined by Eq.~\eqref{equamotinit} are equivalent to those obtained from
\be\label{equamot}
\frac{\delta S}{\delta\Phi^\star}=0~~~~\mbox{and}~~~~\frac{\delta S^\star}{\delta\Phi}=0~.
\ee
This choice places the zero mode in the right eigenspectrum of the non-Hermitian Klein-Gordon operator. The alternative choice 
\be\label{equamotbis}
\frac{\delta S}{\delta\Phi}=0~~~~\mbox{and}~~~~\frac{\delta S^\star}{\delta\Phi^\star}=0
\ee
corresponds to switching the coupling $\mu^2 \ \leftrightarrow - \mu^2$ and choosing the zero mode to lie instead in the left eigenspectrum.
However \cite{Alexandre:2017foi, Alexandre:2017erl}, this does not change the physical observables, since they depend only on $(\pm\mu^2)^2$. We are therefore
free to choose the equations of motion as in Eqs.~\eqref{equamotinit} and \eqref{equamot}. This reflects the fact that, as in the Hermitian case, 
physical observables are invariant under transformations of the 
discrete $\mathbb{Z}_2\times\mathbb{Z}_2$ group, i.e.~we can absorb a change in the sign of $\mu^2$ by an appropriate field redefinition.

We remark that this freedom to choose the defining equations of motion persists in the Hamiltonian formulation. Specifically, the Legendre transform relating the 
Lagrangian and Hamiltonian descriptions is unaffected by the non-Hermiticity of the potential, since the definition of the conjugate momenta is unchanged from the Hermitian case. 
Of Hamilton's equations, only those for the time-derivatives of the conjugate momenta are affected, and we may freely choose to define the equations of motion with respect to
\be
\label{eq:ham1}
\partial_t\Pi^\dagger =\ -\:\frac{\partial \mathcal{H}}{\partial \Phi^{\dag}}~,
\ee
or, alternatively,
\be
\label{eq:ham2}
\partial_t\Pi =\ -\:\frac{\partial \mathcal{H}}{\partial \Phi}\ \neq\ (\partial_t \Pi^\dagger)^{\dagger}~.
\ee
We emphasise that Eqs.~\eqref{eq:ham1} and Eq.~\eqref{eq:ham2} are not related by Hermitian conjugation since the Hamiltonian density $\mathcal{H}\neq \mathcal{H}^{\dagger}$ is not Hermitian 
--- the operations of Hermitian conjugation and derivation with respect to time do not commute (i.e.~$\partial_t\Pi\ne\partial_t^\dagger \Pi$, where the meaning of $\partial_t^\dagger$ 
follows from the Hamilton equations of motion). As in the case of the Euler-Lagrange equations, however, the two choices are related by the 
transformation $\mu^2\to -\mu^2$, under which physical observables remain unchanged.

As discussed in Refs.~\cite{Alexandre:2017foi, Alexandre:2017erl}, the eigenvectors ${\bf e_\pm}$ of the mass matrix, appearing in the Lagrangian 
(\ref{eq:LagrangianFreeScalar}) and corresponding to the eigenvalues $M_\pm^2$, are not orthogonal
with respect to Hermitian conjugation, i.e.~$({\bf e_+})^\dagger\cdot{\bf e_-}\ne0$, but they are orthogonal with respect to $
\mathcal{PT}$ conjugation, i.e.~$({\bf e_+})^\ddagger\cdot{\bf e_-}=0$. The
inner product of states (in flavour space) must therefore be defined with $\mathcal{PT}$-conjugate fields, and the
time evolution of the system is then derived in the usual way by expanding the fields on the basis vectors ${\bf e_\pm}$, 
with the corresponding creation operators evolving with the factor $\exp[itE_\pm(\mathbf{p})]$, where 
$E_\pm^2(\mathbf{p})=\mathbf{p}^2+M_\pm^2$. Note that the canonical equal-time commutation relations, 
for each field $\phi_i$ and its conjugate momentum $\pi_i=\dot{\phi_i}^\star$, are not modified by the non-Hermiticity, in accordance with the discussion 
of the Hamiltonian description above. 
Once the equations of motion are chosen according to either Eqs.~(\ref{equamot}) or Eqs.~(\ref{equamotbis}), quantisation therefore follows the usual steps and, 
as stated in Ref.~\cite{BM}, the Heisenberg picture holds since the Hamiltonian, although non-Hermitian, remains the generator for time evolution.

To elaborate further on the consistency of the choice of equations of motion, it is convenient to define new field variables $(\Xi,\bar{\Xi})$, for which the mass matrix is diagonal:
\be
\label{eq:similarity}
\Xi\ \equiv\ R\Phi\qquad {\rm and}\qquad \bar\Xi\ \equiv\ \Phi^\dagger R^{-1}~,
\ee
where (for $m_1^2>m_2^2$)
\be \label{Rmatrix}
R \ =\ \mathcal{N}\begin{pmatrix}
	\eta &  1 - \sqrt{1 - \eta^2}  \\1 - \sqrt{1 - \eta^2}   & \eta
\end{pmatrix}
\ee
and
\be
\mathcal{N}^{-1}\ \equiv \ \sqrt{2 \eta^2\: -\: 2\: +\: 2\sqrt{1 \:-\: \eta^2}}~.
\ee
Notice that this is a similarity rather than a unitary transformation, and it is defined only away from the exceptional point $\eta= 1$.

In this basis, the Lagrangian in Eq.~\eqref{eq:LagrangianFreeScalar} takes the form
\be
\mathcal{L}\ =\ \bar{\Xi}\begin{pmatrix} -\Box-M_+^2 & 0 \\ 0 & -\Box-M_-^2 \end{pmatrix}\Xi~.
\ee
The variable $\bar\Xi\neq \Xi^{\dag}$ is the $\mathcal{C}'\mathcal{PT}$-conjugate of $\Xi$: $\bar\Xi\ =\ \Xi^{\ddagger}C'$, where the matrix $C'=RPR^{-1}$. 
The $\mathcal{C}'\mathcal{PT}$ conjugation is the involution with respect to which the positive-definite inner product is 
defined for non-Hermitian, $\mathcal{PT}$-symmetric QFTs~\cite{Bender:2002vv}, and it is in terms of these $\mathcal{C}'\mathcal{PT}$-conjugate variables 
that the partition function can be defined consistently, as explained in Ref.~\cite{AEMS1}. 
The equations of motion, consistent with either Eqs.~\eqref{equamot} or \eqref{equamotbis}, follow straightforwardly from the variations
\be
\frac{\delta S}{\delta \bar{\Xi}}\ =\ 0 \qquad {\rm or} \qquad \frac{\delta S}{\delta \Xi}\ =\ 0~,
\ee
which still correspond to two distinct choices.

Returning to the Lagrangian in Eq.~(\ref{eq:LagrangianFreeScalar}), we have invariance under the global phase transformation 
$\Phi \rightarrow\ e^{-i \theta} \Phi$. However, the corresponding Noether current 
\be\label{jalpha+}
j_+^\alpha\ \equiv\ i\left(\phi_1^\star \partial^\alpha \phi_1 - \phi_1 \partial^\alpha \phi_1^\star \right) 
+ i\left( \phi_2^\star \partial^\alpha \phi_2 - \phi_2 \partial^\alpha \phi_2^\star \right)
\ee
is not conserved when the equations of motion, obtained as described above, are applied. On the other hand, the current corresponding to the transformations 
\be \label{eq:ConservedTransformationsScalar}
\Phi\rightarrow e^{-i\theta P}\Phi=\begin{pmatrix}
e^{-i \theta} \phi_1 \\ e^{i \theta} \phi_2
\end{pmatrix}
\ee
{\it is} conserved, and it is given by 
\be \label{eq: ConservedCurrentScalar}
j_-^\alpha\ =\ i\left(\phi_1^\star \partial^\alpha \phi_1 - \phi_1 \partial^\alpha \phi_1^\star \right) 
- i\left( \phi_2^\star \partial^\alpha \phi_2 - \phi_2 \partial^\alpha \phi_2^\star \right)~.
\ee
We note that the transformation (\ref{eq:ConservedTransformationsScalar}) does {\it not} leave the Lagrangian invariant:
$\delta \mathcal{L}\ =\ \frac{\delta S}{\delta \phi_i} \delta \phi_i\neq\ 0$ (see Ref.~\cite{Alexandre:2017foi}). 
Instead, the Lagrangian transforms into 
\be
\mathcal{L}_{\theta} =\ \partial_\alpha \phi_1^\star \partial^\alpha \phi_1 + \partial_\alpha \phi_2^\star \partial^\alpha \phi_2 - m_1^2 |\phi_1|^2 - m_2^2 |\phi_2|^2 
-\mu^2\big( e^{+2i \theta} \phi_1^\star \phi_2 - e^{-2i \theta} \phi_2^\star \phi_1\big)~.
\ee
However, even though the Lagrangian is different from (\ref{eq:LagrangianFreeScalar}), the physical observables remain unchanged and describe the same physical system~\cite{AEMS1}. 
This implies that there is a one-parameter family of non-Hermitian Lagrangians that describe the same physics~\cite{AEMS1}. 
As we will see, however, the situation is quite different in the case of local symmetries.

\section{Spontaneous symmetry breaking and the Goldstone mode}
\label{sec:global}

Before considering the case of local symmetries, we first review how the Goldstone theorem can be extended~\cite{AEMS1} from the standard Hermitian case
to that of a non-Hermitian, $\mathcal{PT}$-symmetric system, the only requirement 
being the existence of a conserved current $j^\alpha$ and a non-trivial vacuum $v$ for which $\varphi (v)\ \neq\ v$,
where $\varphi$ is the transformation corresponding to the current $j^\alpha$.
A simple example of such a non-Hermitian, $\mathcal{PT}$-symmetric system is given by the Lagrangian 
\be \label{eq:lagr}
\mathcal{L} \ =\ \partial_\alpha \phi_1^\star \partial^\alpha \phi_1 + \partial_\alpha \phi_2^\star \partial^\alpha \phi_2 + m_1^2 |\phi_1|^2 - m_2^2 |\phi_2|^2 
- \mu^2 \left( \phi_1^\star \phi_2 - \phi_2^\star \phi_1 \right) - \frac{g}{4} |\phi_1|^4~. 
\ee
Using the equations of motion, we find a non-trivial vacuum that is a solution of the equations 
\begin{subequations}
\bea 
0 \ &=&\ \big(g |\phi_1|^2 - 2m_1^2 \big) \phi_1 + 2\mu^2 \phi_2~, \\
0 \ &=&\ m_2^2 \phi_2 - \mu^2 \phi_1~.
\eea
\end{subequations}
This vacuum is given by 
\be \label{eq:vac}
\begin{pmatrix}
v_1 \\ v_2
\end{pmatrix}\ =\ \sqrt{2\frac{m_1^2 m_2^2 - \mu^4}{g m_2^2}}\begin{pmatrix}
1 \\ \frac{\mu^2}{m_2^2}
\end{pmatrix}~,
\ee
up to an overall complex phase.

The potential for the fluctuations can be written in the form
\bea 
\label{eq:U}
U(\hat{\phi}_1, \hat{\phi}_2) \ &=&\ -\frac{2 \mu^4}{m_2^2} v_1 \hat{\phi}_1 + 2 m_2^2 v_2 \hat{\phi}_2 + {\tilde{m}_1^2}|\hat{\phi}_1|^2 
+ \frac{g}{4}v_1^2 \big( \hat{\phi}_1^2 + (\hat{\phi}_1^\star)^2 \big) \nonumber\\  &+& m_2^2 |\hat{\phi}_2|^2  
+ \mu^2 \big( \hat{\phi}_1^\star \hat{\phi}_2 - \hat{\phi}_2^\star \hat{\phi}_1 \big) + \frac{g}{2}v_1\big (\hat{\phi}_1 + \hat{\phi}_1^\star\big)|\phi_1|^2 + \frac{g}{4}|\phi_1|^4~,
\eea
where {$\tilde{m}_1^2\equiv g v_1^2 - m_1^2$ and we have shifted the fields:} $\phi_1 \equiv v_1 + \hat{\phi}_1$ and $\phi_2 \equiv v_2 + \hat{\phi}_2$.
The linear terms in this potential are a consequence of our non-Hermitian behaviour. Note that they are not symmetric under $\mathcal{PT}$, such that the non-trivial vacuum 
breaks the $\mathcal{PT}$ symmetry of the action. Even so, there remains a region of parameter space in which the eigenspectrum of the fluctuations is real and positive semi-definite
and, though present in the potential, they play no role in the equations of motion, nor their complex conjugates, which are given by
{\begin{subequations}
\bea 
\big(-\Box -\tilde{m}_1^2\big)\hat{\phi}_1\ &=&\ + \mu^2 \hat{\phi}_2  + \frac{g}{2} v_1^2 \hat{\phi}_1^\star 
+ \frac{g}{2} \big( v_1 \hat{\phi}_1^2 + 2 v_1 |\hat{\phi}_1|^2 + |\hat{\phi}_1|^2 \hat{\phi}_1 \big) ~, \\
\big(-\Box -m_2^2\big)\hat{\phi}_2\ &=&\  - \mu^2 \hat{\phi}_1~.
\eea
\end{subequations}}

The mass squared matrix is given by the linear terms in these equations and takes the form 
\be 
M^2\ =\ \begin{pmatrix}
{\tilde{m}_1^2} & \frac{g}{2} v_1^2 & \mu^2 & 0 \\ 
\frac{g}{2} v_1^2 & {\tilde{m}_1^2} & 0 & \mu^2 \\
-\mu^2 & 0 & m_2^2 & 0 \\ 
0 & -\mu^2 & 0 & m_2^2 
\end{pmatrix} ~.
\ee
This matrix has an eigenmode 
\be 
G_1 = \sqrt{\frac{2m_2^4}{m_2^4 - \mu^4}} \Bigg[ \im \big( \hat{\phi}_1 \big) - \frac{\mu^2}{m_2^2} \im\big( \hat{\phi}_2 \big) \Bigg]~,
\label{G1}
\ee
with eigenvalue $\lambda_1 = 0$, which is the Goldstone mode in this model.\footnote{Notice that the normalisation of the Goldstone mode 
(with respect to $\mathcal{PT}$ conjugation) diverges in the limit $\mu^2=\pm m_2^2$ (see the note added).}
We gave in Ref.~\cite{AEMS1} a general proof that such a mode must appear whenever there is a non-trivial vacuum
for which $\varphi(v)\ne v$ holds and verified the persistence of the Goldstone mode (\ref{G1}) at the one-loop level.

The other modes of this model have eigenvalues 
\begin{subequations}
\label{eq:eigenmodes}
\bea
\lambda_2 \ &=&\ m_2^2 - \frac{\mu^4}{m_2^2} ~, \\
\lambda_\pm\ &=&\ \frac{1}{2m_2^2} \Big(2m_1^2 m_2^2 - 3 \mu^4 + m_2^4 \pm \sqrt{\left( 2 m_1^2 m_2^2 - 3\mu^4 - m_2^4\right)^2 - 4 \mu^4 m_2^4} \Big) ~, 
\eea  
\end{subequations}
and are given by 
\begin{subequations}
\bea 
G_2\ &=&\ \sqrt{\frac{2 m_2^4}{m_2^4 - \mu^4}} \Bigg[ \im \big( \hat{\phi}_2 \big) - \frac{\mu^2}{m_2^2} \im \big( \hat{\phi}_1 \big) \Bigg] ~, \\
G_\pm\ &=&\ \frac{\sqrt2}{\sqrt{(\lambda_\pm-m_2^2)^2-\mu^4}}\left[(\lambda_\pm-m_2^2)\mbox{Re}\big(\hat\phi_1\big)+\mu^2\mbox{Re}\big(\hat\phi_2\big)\right]~,
\eea
\end{subequations}
respectively. We note that the masses of these physical modes depend in different ways on the mass parameter $\mu$ that characterises the amount
of non-Hermiticity in the Lagrangian~\eqref{eq:lagr}.

\section{Gauging the $\mathcal{PT}$-symmetric model}
\label{sec:gauged}

\subsection{Naive approach}

We may seek to promote the above global transformations to local transformations by introducing a gauge field $A^\alpha$ and minimally 
coupling it to the scalar fields via the gauge covariant derivatives. 
For the Maxwell equations to have the usual canonical form though, $\partial_\alpha F^{\alpha\beta}=j_{A,-}^\beta$,
we must couple the gauge field to a conserved current with $\partial_\beta j^\beta_{A,-}=0$, since $\partial_\alpha\partial_\beta F^{\alpha\beta}=0$ identically. 
The Lagrangian then takes the form
\be \label{eq:gaugedLagnaive}
\mathcal{L} = [ D_\alpha^+ \phi_1 ]^\star D^\alpha_+ \phi_1 + [D_\alpha^- \phi_2 ]^\star D^\alpha_- \phi_2 - m_1^2 |\phi_1|^2 - m_2^2 |\phi_2|^2 
- \mu^2 \left( \phi_1^\star \phi_2 - \phi_2^\star \phi_1 \right) - \frac{1}{4}F_{\alpha\beta} F^{\alpha\beta}~,
\ee
where the covariant derivatives are $D_\pm^\alpha  =  \partial^\alpha \pm iqA^\alpha$.
The conserved current is
\be 
\label{eq:current}
j^\alpha_{A,-} =\ {i q} \big( \phi_1^\star D_+^\alpha \phi_1 - \phi_1 [D_+^\alpha \phi_1]^\star \big) 
- {i q}\big( \phi_2^\star D_-^\alpha \phi_2 - \phi_2 [D_-^\alpha \phi_2]^\star \big)~,
\ee
and the kinetic terms in the Lagrangian are invariant under the transformations 
\begin{subequations}
\bea
\phi_1(x) & \rightarrow & \phi_1(x) e^{-i q f(x)}~,\\
\phi_2(x) & \rightarrow & \phi_2(x) e^{+iq f(x)}~,\\
A^\alpha(x) & \rightarrow & A^\alpha(x) + \partial^\alpha f(x)~.
\eea
\end{subequations}
The kinetic term could also be written in terms of $\mathcal{D}_\alpha\Phi$ with $\mathcal{D}_\alpha=\mathbb{I}_2\partial_\alpha+iqPA_\alpha$, 
making manifest the role played by the parity matrix $P$ in the definition of the conserved current.

However, with this form of coupling, we see that the non-Hermitian mass term explicitly breaks gauge invariance. Specifically, the gauge transformation yields a local mass squared matrix
\begin{equation}
\label{eq:Mloc}
M^2(x)\ =\ \begin{pmatrix} m_1^2 & \mu^2e^{+2iqf(x)} \\ -\:\mu^2e^{-2iqf(x)} & m_2^2 \end{pmatrix} 
\equiv \begin{pmatrix}m_1^2 & \tilde{\mu}^2(x) \\ \left[ - \tilde{\mu}^2(x) \right]^\star & m_2^2
\end{pmatrix}~.
\end{equation}
The eigenspectrum is unaffected by the additional phases in the off-diagonal elements of Eq.~\eqref{eq:Mloc}, 
and the squared mass eigenvalues remain real and independent of the gauge function $f$, since they involve $\tilde\mu^2(x)[\tilde\mu^2(x)]^\star=\mu^4$. 
Rotating to the mass eigenbasis via the similarity transformation in Eq.~\eqref{eq:similarity}, the gauge dependence is shifted to the gauge interactions, 
since the matrix $R$, which is modified to the local form
\be
R(x) \ =\ \mathcal{N}\begin{pmatrix}
	\eta\,e^{-\,2iqf(x)} &  1 - \sqrt{1 - \eta^2}  \\1 - \sqrt{1 - \eta^2}   & \eta\,e^{+2iqf(x)}
\end{pmatrix}\;,
\ee
does not commute with the $P$ matrix appearing in the gauge coupling, i.e.~$R^{-1}PR\neq P$. As a result, and while the eigenspectrum is gauge invariant, 
we find that the photon acquires a mass beyond tree-level; namely, at the one-loop level, we find that the polarisation tensor is not transverse:
\be
k_\alpha\Pi^{\alpha\beta}(k^2=0)=\frac{q^2}{8\pi^2}\,\frac{k^\beta\mu^4}{(M_+^2-M_-^2)^3}\bigg[M_+^4-M_-^4+2M_+^2M_-^2\ln\bigg(\frac{M_-^2}{M_+^2}\bigg)\bigg]\;.
\ee

The above observations indicate that the non-Hermitian deformation of massless gauge theories is problematic, due to the necessary violation of gauge invariance. 

One could modify the naive Lagrangian (\ref{eq:gaugedLagnaive}) though, if one wishes to maintain a coupling to the conserved current as well as gauge invariance. One 
might be tempted to introduce a non-minimal coupling, with the Lagrangian
\bea
\mathcal{L}_W &=& [D_\alpha^+\phi_1]^\star D_+^\alpha\phi_1+[D_\alpha^-\phi_2]^\star D^\alpha_-\phi_2-m_1^2|\phi_1|^2-m_2^2|\phi_2|^2\nonumber\\
&&-\mu^2\Big(W^{\star2}(x)\phi_1^\star\phi_2-W^2(x)\phi_2^\star\phi_1\Big) - \frac{1}{4}F_{\alpha\beta} F^{\alpha\beta}~,
\eea
where
\begin{equation}
W(x)\ =\ \exp\bigg[iq\int^x A_\alpha{\rm d}y^\alpha\bigg]
\end{equation}
is a Wilson line \cite{Wilson:1974sk}, running along a path from the boundary (at infinity) to the spacetime point $x$. Under a gauge transformation (chosen to vanish at infinity), we have
\begin{equation}
W(x)\ =W(x)e^{iqf(x)}\;,
\end{equation}
and the Lagrangian is invariant. However, we have traded the problem of gauge invariance for the path-dependence of the Wilson line. 
Moreover, we see that the gauge field now couples to the non-Hermitian term, such that the equation of motion for the gauge field obtains an imaginary part, 
potentially violating the reality of the gauge field.

\subsection{Modification of charge allocation}

In order to keep gauge invariance, we can instead couple the gauge field to the \emph{non-conserved} current
\be 
\label{eq:current2}
j^\alpha_{A,+} =\ {i q} \big( \phi_1^\star D^\alpha \phi_1 - \phi_1 [D^\alpha \phi_1]^\star \big) 
+ {i q}\big( \phi_2^\star D^\alpha \phi_2 - \phi_2 [D^\alpha \phi_2]^\star \big)~,
\ee
where $D^\alpha  =  \partial^\alpha + iqA^\alpha$, with divergence 
\be
\partial_\alpha j^\alpha_{A,+}=2iq\mu^2(\phi_2^\star\phi_1-\phi_1^\star\phi_2)~.
\ee
In this case, $\phi_1$ and $\phi_2$ are assigned identical charges, and the non-Hermitian mass term is gauge invariant. 
However, in order to ensure that the Maxwell equations are consistent, since $\partial_\beta j^\beta_{A,+}\neq 0$,
we need to add to the Lagrangian the term
\be
-\frac{1}{2\xi}(\partial_\alpha A^\alpha)^2~,
\ee
which would, in the Hermitian case, correspond to fixing a covariant gauge that satisfies the Lorenz gauge condition $\partial_\alpha A^\alpha=0$. 
Notice that, with the addition of this term, and as in the Hermitian case, the gauge functions must satisfy the constraint $\Box f=0$, such that we only have a restricted gauge invariance.

The equation of motion for the gauge field becomes
\be\label{modifMaxwell}
\Box A^\alpha-(1-1/\xi)\partial^\alpha\partial_\beta A^\beta=j^\alpha_{A,+}~,
\ee
and its divergence yields
\be
\label{eq:constraint}
\frac{1}{\xi}\Box\partial_\alpha A^\alpha\ =\ 2iq\mu^2(\phi_2^\star\phi_1-\phi_1^\star\phi_2)~.
\ee
We see that the non-Hermiticity precludes the Lorenz gauge condition,
and the consistency of the Maxwell equation instead leads to the constraint
\be
\Box\pi_0\ =\ 2iq\mu^2(\phi_1^\star\phi_2-\phi_2^\star\phi_1)~,
\ee
where $\pi_0=-\,\partial_\alpha A^\alpha/\xi$ is the momentum conjugate to $A_0$.

As a last remark, we note that the above formulation arises naturally from the St\"uckelberg mechanism \cite{Stueckelberg:1900zz} (see, e.g., Ref.~\cite{Ruegg:2003ps}), 
in the limit where the vector mass goes to zero. To see this, we introduce
an extra real scalar field $\rho$, and consider the Lagrangian
\bea \label{eq:ProcaStuck}
\mathcal{L}_\rho & =& [ D_\alpha \phi_1 ]^\star D^\alpha \phi_1 + [D_\alpha \phi_2 ]^\star D^\alpha \phi_2 - m_1^2 |\phi_1|^2 - m_2^2 |\phi_2|^2 
- \mu^2 \left( \phi_1^\star \phi_2 - \phi_2^\star \phi_1 \right)\nonumber\\
& \phantom{=}&- \frac{1}{4}F_{\alpha\beta} F^{\alpha\beta}+\frac{1}{2}\,\big(m_0A_\alpha-\partial_\alpha\rho\big)\big(m_0A^\alpha-\partial^\alpha\rho\big)
-\frac{1}{2\xi}\big(\partial_\alpha A^\alpha+\xi m_0\rho\big)^2~.
\eea
This Lagrangian is invariant under the gauge transformations
\begin{subequations}
\bea
\phi_{1,2}(x) & \rightarrow & \phi_{1,2}(x) e^{-i q f(x)}~,\\
A^\alpha(x) & \rightarrow & A^\alpha(x) + \partial^\alpha f(x)~,\\
\rho(x) & \rightarrow & \rho(x) + m_0 f(x)~,
\eea
\end{subequations}
where the gauge function satisfies $(\Box+\xi m_0^2)f=0$. The equation of motion for $A_\alpha$ then yields Eq.~\eqref{modifMaxwell} in the limit $m_0\to0$, where the scalar 
$\rho$ decouples from the system, and the constraint \eqref{eq:constraint} necessarily arises.

\subsection{Reality of the background gauge field}

We discuss here the reality of the background gauge field $A^\alpha_b$ after quantum corrections. $A_b^\alpha$ is defined as
\be
A^\alpha_b=\frac{1}{Z}\frac{\delta Z}{\delta J_\alpha}~,
\ee
where $Z$ is the Euclidean partition function and $J_\alpha$ is the corresponding source. $Z$ is ${\cal PT}$-symmetric and can defined as
\be
Z=\int{\cal D}[A_\alpha,\Phi,\Phi^\ddagger]
\exp\left(-S_E+\int{\rm d}^4x\; \Big(J_\alpha A^\alpha+\chi_1^\mathcal{PT}\phi_1+\phi_1^\mathcal{PT}\chi_1+\chi_2^\mathcal{PT}\phi_2+\phi_2^\mathcal{PT}\chi_2\Big)\right)~,
\ee
where $\chi_k$ and $\chi_k^\mathcal{PT}$ are the sources for $\phi_k^\mathcal{PT}$ and $\phi_k$, respectively.

For $A_b^\alpha$ to be real, it is enough to find a condition for the Euclidean partition function to be real, although the Euclidean action $S_E$ has an imaginary part, 
which is opposite in sign to ${\rm Im}\,S$, given in Eq.~\eqref{imaginary}. 
This condition can be achieved by choosing the transformation of the sources $\chi_k$ under $\mathcal{PT}$ appropriately.
For this, we note that the partition function can also be written
\be
Z=\int{\cal D}[A_\alpha,\Phi,\Phi^\ddagger]
\exp\left(-S_E+\int{\rm d}^4x\; \Big( J_\alpha A^\alpha+\chi_1^\mathcal{PT}\phi_1+\phi_1^\star\chi_1+\chi_2^\mathcal{PT}\phi_2-\phi_2^\star\chi_2\Big)\right)~,
\ee
such that
\be
Z^\star=\int{\cal D}[A_\alpha,\Phi,\Phi^\ddagger]\exp\left(-S_E^\star+\int{\rm d}^4x\; \Big(J_\alpha A^\alpha+(\chi_1^\mathcal{PT})^\star\phi_1^\star+\phi_1\chi_1^\star
+(\chi_2^\mathcal{PT})^\star\phi_2^\star-\phi_2\chi_2^\star\Big)\right)~,
\ee
which, after the change of variable $\phi_2\to-\phi_2$, leads to
\be
Z^\star=\int{\cal D}[A_\alpha,\Phi,\Phi^\ddagger]\exp\left(-S_E+\int{\rm d}^4x\; \Big(J_\alpha A^\alpha+(\chi_1^\mathcal{PT})^\star\phi_1^\star+\phi_1\chi_1^\star
-(\chi_2^\mathcal{PT})^\star\phi_2^\star+\phi_2\chi_2^\star\Big)\right)~.
\ee
Imposing $Z^\star=Z$ implies then $\chi_1^\mathcal{PT}=\chi_1^\star$ and $\chi_2^\mathcal{PT}=\chi_2^\star$.
Note that this is consistent with the $\mathcal{PT}$ properties of the scalar background field $\phi_2^b$, defined as
\be
\phi_2^b=\frac{1}{Z}\frac{\delta Z}{\delta \chi_2^\mathcal{PT}}~,
\ee
since
\be
(\phi_2^b)^\mathcal{PT}=\frac{1}{Z}\frac{\delta Z}{\delta \chi_2}=-(\phi_2^b)^\star~.
\ee
As a consequence, $\mathcal{PT}$ symmetry ensures that the gauge field remains real after quantum corrections, even though it is coupled to a non-Hermitian scalar sector.

Finally, one can also conclude from the reality of the partition function that physical observables depend on $\mu^4$ only. Indeed, for $Z$ to be real, the imaginary part of 
the action, cf.~Eq.~\eqref{imaginary}, 
must contribute to the calculation of $Z$ with even powers, and thus with $(\pm\mu^2)^2$. This property, predicted at the tree level, can thus be extended to the full quantum system.

\section{Englert-Brout-Higgs mechanism}
\label{sec:gauge}

In this section, we show that a gauge-invariant mass can be generated at tree-level by the Englert-Brout-Higgs mechanism. 
Given the considerations in Sec.~\ref{sec:gauged}, we consider the Lagrangian 
\bea \label{eq:gaugedLag}
\mathcal{L} &=&\ [ D_\alpha \phi_1 ]^\star D^\alpha \phi_1 + [D_\alpha\phi_2 ]^\star D^\alpha \phi_2 + m_1^2 |\phi_1|^2 - m_2^2 |\phi_2|^2 
- \mu^2 \left( \phi_1^\star \phi_2 - \phi_2^\star \phi_1 \right)\nonumber\\&& \qquad - \frac{g}{4}|\phi_1|^4 - \frac{1}{4}F_{\alpha\beta} F^{\alpha\beta}
- \frac{1}{2\xi}\left( \partial_\alpha A^\alpha \right)^2~,
\eea
where we emphasise that the would-be gauge-fixing term $-(\partial_\alpha A^\alpha)^2/(2\xi)$ is necessary for consistency of the model.

The vacuum expectation value for the scalar fields is the same as in the global model (\ref{eq:vac}), and
we can express the Lagrangian (\ref{eq:gaugedLag}) in terms of the shifted fields:
\bea 
\label{shiftedL}
\mathcal{L} \ &=&\ \partial_\alpha \hat{\phi}_1^\star \partial^\alpha \hat{\phi}_1 + \partial_\alpha \hat{\phi}_2^\star \partial^\alpha \hat{\phi}_2 
-U(\hat{\phi}_1,\hat{\phi}_2)
- \frac{1}{4} F_{\alpha\beta}F^{\alpha\beta}  - \frac{1}{2\xi}\left( \partial_\alpha A^\alpha \right)^2 
\nonumber\\&&\qquad+ q^2 A_\alpha A^\alpha \big( |v_1 + \hat{\phi}_1|^2 + |v_2 + \hat{\phi}_2|^2\big) -A_\alpha j_+^\alpha~,
\eea   
where $U(\hat{\phi}_1,\hat{\phi}_2)$ is defined in Eq.~\eqref{eq:U} and $j_+^\alpha$ is the current in Eq.~\eqref{jalpha+}.
We then obtain the equations of motion
\begin{subequations}
\label{EofMs}
\begin{align}
\big(-D^2    -\tilde{m}_1^2\big) \hat{\phi}_1
&=  +\mu^2 \hat{\phi}_2-q^2v_1A^2 +iqv_1\partial_\alpha A^\alpha\nonumber\\
&+ \frac{g}{2} v_1^2\hat{\phi}_1^\star  + \frac{g}{2}\big(v_1 \hat{\phi}_1^2
+ 2v_1 |\hat{\phi}_1|^2+|\hat{\phi}_1|^2 \hat{\phi}_1\big) ~,  \\
\big(-D^2 -m_2^2\big) \hat{\phi}_2 &=  -\mu^2 \hat{\phi}_1 - q^2 v_2A^2+iqv_2\partial_\alpha A^\alpha~,  \\
\big(-\Box -M_A^2\big)A^\alpha+(1-1/\xi)\partial^\alpha\partial_\beta A^\beta &= 
2q^2\big( v_1^{\star}\hat{\phi}_1+v_1\hat{\phi}_1^{\star}+ v_2^{\star}\hat{\phi}_2+v_2\hat{\phi}_2^{\star}\big)A^\alpha\nonumber\\
&+2q^2\big(|\hat{\phi}_1|^2+|\hat{\phi}_2|^2\big)A^\alpha -j_+^\alpha~,
\end{align}
\end{subequations}
where
\be 
M_A^2 = 2q^2 \left( |v_1|^2 + |v_2|^2 \right)
\ee
is the gauge-invariant squared-mass of the gauge boson. Therefore, although the non-Hermitian model has non-trivial features related to gauge invariance, 
the usual Englert-Brout-Higgs mechanism still holds.

\section{Conclusions and perspectives}
\label{sec:conx}

We have shown in this paper how the Englert-Brout-Higgs mechanism~\cite{Englert:1964et,Higgs:1964pj} for generating masses for gauge bosons can be generalised
from the familiar case of Hermitian QFTs to the more general framework of $\mathcal{PT}$-symmetric field theories. 
However, we have seen that to preserve gauge invariance in the non-Hermitian gauge theories described here, it is necessary to couple the gauge field to the non-conserved current. 
The consistency of the Maxwell equations then requires the inclusion of the would-be gauge fixing term but precludes the Lorenz gauge and leads to a particular constraint on 
the gauge field that depends on the non-Hermitian structure of the theory.

We have restricted our attention in this work to the Abelian case, and it would clearly be interesting to explore the possible
extension to the non-Abelian case~\cite{Kibble}, which will require a careful re-examination of the quantisation procedure
for non-Abelian gauge fields in the context of $\mathcal{PT}$-symmetric field theories. Such an analysis should be completed
by a study of renormalisation and unitarity, including the possibility of non-Hermitian gauge anomalies.
We note that the scalar fields in the $\mathcal{PT}$-symmetric model we have studied could in principle
be elevated to doublets of an SU(2) gauge group so, if these issues can be resolved, one might consider
using this model as the basis for the possible construction of a non-Hermitian
extension of the Standard Model, as well as other new scenarios in particle modeling that might also incorporate 
non-Hermitian extensions of the Yukawa sector~\cite{Alexandre:2015kra,Alexandre:2017fpq}.\\

\noindent
{\bf Note added:}  
While this work was being prepared, we saw Ref.~\cite{Mannheim}, in which Goldstone bosons and the Englert-Brout-Higgs
mechanism in non-Hermitian theories are discussed from a complementary perspective. We thank Philip Mannheim for kindly drawing
our attention to his interesting paper. This also made us aware of a consistent error in the normalisation of the Goldstone modes in our previous work~\cite{AEMS1} 
(corrected herein), which obscured the behaviour of the exceptional point $\mu^2=\pm m_2^2$, as discussed in detail in Ref.~\cite{Mannheim}.

\section*{Acknowledgements}

We thank the Referee for pointing out the need to address ambiguities related to gauge invariance.
The work of JA and JE was supported by the United Kingdom STFC Grant ST/P000258/1, and that of JE also by the Estonian Research Council via a Mobilitas Pluss grant.  
The work of PM was supported by a Leverhulme Trust Leadership Award.

\end{document}